\documentclass[graybox]{svmult}

\usepackage{mathptmx}       
\usepackage{helvet}         
\usepackage{courier}        
\usepackage{type1cm}        
\usepackage{makeidx}         
\usepackage{graphicx}        
\usepackage{multicol}        
\usepackage[bottom]{footmisc}

\makeindex             

\def\lesssim{\mathrel{\hbox{\rlap{\hbox{\lower4pt\hbox{$\sim$}}}\hbox{$<$}}}}
\def\gtrsim{\mathrel{\hbox{\rlap{\hbox{\lower4pt\hbox{$\sim$}}}\hbox{$>$}}}}

\begin{document}

\title*{A Semi-automatic Search for Giant Radio Galaxy Candidates and their Radio-Optical Follow-up}
\author{I.\ del C.\ Santiago-Bautista, C.\,A. Rodr\'{i}guez, H.\ Andernach, R.\ Coziol, \\
   J.\,P.\ Torres-Papaqui, E.\,F. Jim\'enez Andrade, I. Plauchu-Frayn, E. Momjian}
\authorrunning{Radio-Optical Follow-up of Giant Radio Galaxy Candidates}
\institute{I.\ del C.\ Santiago-Bautista, C.\,A. Rodr\'{i}guez, H.\ Andernach, R.\ Coziol,
   J.\,P.\ Torres-Papaqui \at Deptartamento de Astronom\'{i}a, DCNE, Universidad de Guanajuato, Guanajuato, Mexico \\
  \email{isantiago,carlos,heinz,rcoziol,papaqui@astro.ugto.mx}
\and E.\,F. Jim\'enez Andrade \at Inst.\ Nac.\ de Astrof\'{i}ca, \'Optica y Electr\'onica, Tonantzintla, Pue., Mexico
   \email{ericja@inaoep.mx}
\and I.~Plauchu-Frayn \at  Instituto de Astronom\'{i}a, UNAM, Ensenada, B.C., Mexico
    \email{ilse@astrosen.unam.mx}
\and E.~Momjian \at NRAO, Socorro, NM 87801, USA  \email{emomjian@nrao.edu}
}
%
%
\maketitle

\abstract{
We present results of a search for giant radio galaxies (GRGs) with 
a projected largest linear size in excess of 1\,Mpc. 
We designed a computational algorithm to identify contiguous emission regions, 
large and elongated enough to serve as GRG candidates, and applied it 
to the entire 1.4-GHz NRAO VLA Sky survey (NVSS). In a subsequent visual inspection 
of 1000 such regions we discovered 15 new GRGs, as well as many other candidate 
GRGs, some of them previously reported, for which no redshift was known.
Our follow-up spectroscopy of 25 of the brighter hosts using two 2.1-m telescopes 
in Mexico, and four fainter hosts with the 10.4-m Gran Telescopio Canarias (GTC),
yielded another 24 GRGs.
We also obtained higher-resolution radio images with the Karl G.\ Jansky Very 
Large Array for GRG candidates with inconclusive radio structures in NVSS.
}

\section{Introduction}
\label{sec:1}

Giant Radio Galaxies (GRGs) are defined in the literature as having
extended radio emission over a (projected) largest linear size (LLS) 
$>$1\,Mpc, using  values of H$_0$\,=\,50--100 km\,s$^{-1}$\,Mpc$^{-1}$.
Those are very rare objects. Indeed, from many articles published prior to 
our study (e.g., \cite{ishw99,lara01,schoenm01,mach06}), homogenising the data
to H$_0$\,=\,75~km\,s$^{-1}$\,Mpc$^{-1}$, we compiled a list of only 100 GRGs.

Statistical analyses of samples of GRGs \cite{ishw99,kompas09} suggest
that their extreme sizes neither can be explained by a preferred
orientation in the plane of the sky, nor by a location in less dense
regions of the Universe, nor by more powerful jets feeding their
lobes and thus reaching further out in intergalactic space. Instead,
\cite{kompas09} argued that it is an exceptionally long-lasting radio
activity in $\sim$10\,\% of FR\,II sources \cite{fr74} that allows GRGs
to develop. On the other hand, \cite{malare15} found evidence for the
lobes of GRGs to be oriented normal to the major axes of galaxy overdensities
near the hosts.  Thus, the reason why some radio galaxies become giants
is still not fully understood, and larger samples of GRGs are desirable
to clarify this issue.

Many new GRGs were recently discovered by us \cite{and12} from a visual
inspection of large-area radio surveys and subsequent identification 
of the host, e.g., in NED \cite{ned} or the SDSS \cite{alam15}. 
In order to increase our discovery rate we designed
a computational algorithm that can be applied directly to radio survey images.

\section{Automated Search in the NVSS Image Atlas}
\label{sec:2}

The NVSS at 1.4\,GHz \cite{condon98} currently provides the best combination of 
sensitivity to radio sources of large angular size ($\lesssim16'$), 
low brightness (1-$\sigma\sim$0.45\,mJy\,beam$^{-1}$), angular resolution 
(45$''$), and coverage ($\delta_{2000}>-40^{\circ}$ or 82\,\% of the sky). 
The NVSS image atlas contains 2326 images of 4$^{\circ}\times4^{\circ}$
with 1024$^2$ pixels of 15$''\times15''$.
To detect new GRGs in these images, we designed an algorithm to find
contiguous emission regions, large and elongated enough to suggest the 
presence of a GRG.

First, the images were binarized by setting all pixels above 3 times the noise level
to 1, and all others to 0.
Then we applied the {\it closing} procedure, which consists of two steps: 
1) the {\it erosion} operator sets a pixel to zero if any of its 8 neighbors
is zero; 2) the {\it dilatation} operator sets a pixel to 1 if any
of its 8 neighbor pixels has value~1. Thus, {\it closing} provides an
image cleaned from noise pixels. After that, we perform {\it region growing}, 
which selects only contiguous regions of pixels of value~1 that are 
larger than a minimum number of pixels. To avoid spurious detections,
we also excluded from our search those regions with noise levels 
$\gtrsim$0.6\,mJy\,beam$^{-1}$, e.g.\ near strong sources or close to the 
Galactic plane, as explained in \cite{sil12msc}.

Limiting the region size to at least 300 pixels or 18\,arcmin$^2$
we obtained 1000 such regions. Since our regions were chosen to be 
contiguous, and many radio galaxies are known to show two or three 
separate, neighboring emission regions (core and lobes), we used visual
inspection to detect these cases. To find the host galaxy, we overlaid 
NVSS contours with optical images from DSS \cite{dss} or SDSS \cite{alam15}.
When available, we used FIRST images \cite{bwh95} 
to look for faint ($\lesssim$2\,mJy) radio cores between widely spaced radio lobes in
NVSS. We found optical hosts for 160 candidates with redshifts in NED.
Of these, 15 turned out to be previously unreported GRGs. 
For many of the remaining candidates, plus several of those found
in \cite{and12}, we retrieved photometric redshifts from SDSS \cite{alam15} 
or \cite{bilicki14,brescia14}. This allowed us to estimate their linear 
sizes, and select the largest sources with sufficiently bright host galaxies
for optical spectroscopy.

For many of our GRG candidates the angular resolution of NVSS is not
sufficient to decide on the optical host galaxy, often because the lobes
are far apart and no central compact source is detected that would
indicate the host. For some of these we obtained Karl G.\ Jansky Very Large
Array (VLA) observations at higher angular resolution and/or frequency.

\section{Follow-up with Optical Spectroscopy and Radio Imaging}
\label{sec:3}

From the list of candidates for which an estimation of their linear size
was possible, we selected those that could be larger than $\sim$0.7\,Mpc.
For the optically brighter hosts ($r\lesssim$16.5\,mag) we obtained spectroscopy
with two 2.1-m telescopes in northern Mexico: the Obs.\ Astron\'omico Nacional
(OAN, San Pedro M\'artir) during several runs in 2013 and 2014, and the
Obs.\ Astrof\'{i}sico G.\,Haro (OAGH, Cananea) in April 2014. Spectra of
four fainter hosts ($r\gtrsim$18\,mag) were obtained with the
OSIRIS instrument on the 10.4-m Gran Telescopio Canarias (GTC) in Spain.

With our spectroscopy at the 2.1-m telescopes we were able to confirm
18 GRGs with LLS$>$1\,Mpc, and several more with smaller radio sizes.
Of the four candidates observed with GTC, three have sizes of 1.2--1.5\,Mpc,
and one has LLS$\sim$0.8\,Mpc. 

For another 14 very extended sources, with either undetected radio core or
uncertain radio structure, we obtained VLA observations in C-configuration
at higher angular resolution than NVSS. Two of the most clear-cut
results are shown in~Fig.~1.

\begin{figure}[h!]
\includegraphics[scale=.34]{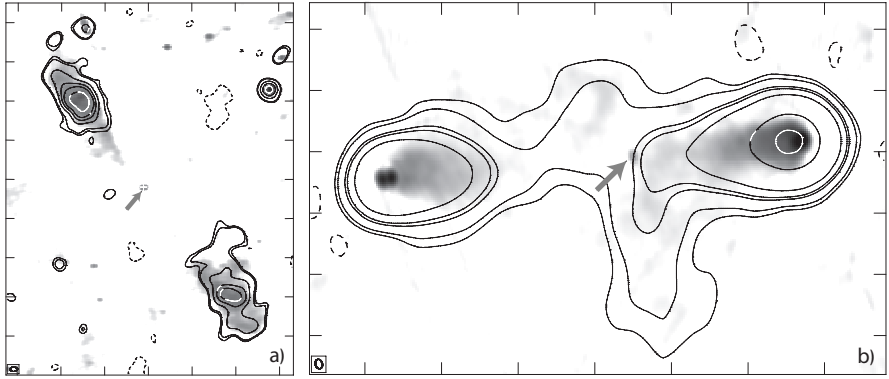}
%
%
\caption{Two of our VLA follow-up observations in grey-scale, with a noise of
0.1\,mJy\,beam$^{-1}$. Contours are from NVSS,
starting at 1.3\,mJy\,beam$^{-1}$ (3\,$\sigma$).~
(a)~~J0047+5339: NVSS shows two amorphous lobes, with no obvious radio core. 
Our new VLA L-band (1--2\,GHz) image with 21$''\times14''$ ($\alpha\times\delta$)
resolution reveals 
collimated structures in the lobes for the first time. We also detect a 0.8-mJy
core of an R=16.3\,mag host galaxy, for which we obtained a redshift of 0.146 at OAN.
The source's angular size of $\sim16'$ implies an LLS of 2.3\,Mpc.~
(b)~J0157+0209: The NVSS image only shows diffuse emission and no radio core.
Our new VLA S-band (2--4\,GHz) image, with 7$''\times10''$ ($\alpha\times\delta$)
resolution, clearly detects
a 1.8-mJy core, coincident with a host of $r\,'=$17.8\,mag at $z=0.2217$ \cite{alam15}. 
The source's angular size of 7\,$'$ thus corresponds to an LLS of 1.4\,Mpc.
}
\label{fig:1}       
\end{figure}

\section{Conclusions}
\label{sec:4}

We developed and applied a semi-automatic method to find GRGs in 
the NVSS radio survey. Radio-optical overlays allowed us to find 
the host galaxy and redshift, resulting in 15 previously unreported 
GRGs. For another 29 candidates without known spectroscopic redshift,
we obtained spectra with 2-m and 10-m class telescopes. 
Of these, we confirmed 24 new GRGs with sizes from 1.0 to 2.3 Mpc 
and several others of smaller size. So far, our project has 
uncovered 39 new GRGs.

Most of our spectroscopic redshifts fall within $\pm$15\,\% of the photometric
ones given in \cite{alam15,bilicki14,brescia14}. We
also found many GRG candidates with $|b|<10^{\circ}$, for which the host
galaxy is obvious, relatively bright, and little affected by Galactic
extinction.

We also obtained new radio observations with the VLA of 14 GRG candidates,
in order to find the radio nucleus or confirm the physical connection
of the radio structure, and to estimate the dynamical ages of the lobes.
Analysis of these data is in progress.

Our optical spectroscopy is ongoing at OAN and GTC during 2015. In our 
GTC observations we are aiming at spectroscopy of GRG candidates at 
z$>$0.5 and LLS$>$2.5\,Mpc to probe the density and physical conditions of GRGs
in the intermediate-redshift Universe. 

\begin{acknowledgement}
HA, CARR, JPTP \& ISB were supported by DAIP-UG grant \#318/13, and ISB by travel grants
from CONACyT and Univ.\ de Guanajuato. 
EFJA is grateful to Academia Mexicana de Ciencias (AMC) for a summer student grant.
We appreciate the help of R.\,F.\ Maldonado S\'anchez in the inspection of
radio survey images. This work is based upon observations acquired at
Obs.\ Astron.\ Nacional, San Pedro M\'artir (OAN-SPM), B.C., Mexico,
at the Obs.\,Astrof\'{i}sico G.\,Haro (OAGH), Cananea, Son., Mexico, and with
the Gran Telescopio Canarias, La~Palma, Spain.
The National Radio Astronomy Observatory is a facility of the National
Science Foundation operated under cooperative agreement by Associated
Universities, Inc.
\end{acknowledgement}

\end{document}